\documentclass[12pt]{article}
\usepackage{amssymb,amsmath}
\usepackage{epsfig}
\newcommand{\sect}[1]{\setcounter{equation}{0}\section{#1}}

\def\N{{\mathcal N}}
\def\L{{\mathcal L}}

\def\a{\alpha}

\def\d{\delta}
\def\g{\gamma}

\def\s{\sigma}
\def\t{\tau}
\def\x{\xi}

\def\m{\mu}
\def\n{\nu}
\def\k{\kappa}
\def\f{\phi}
\def\e{\eta}

\def\l{\lambda}

\def\p{\partial}
\def\rb{\right}
\def\lb{\left}

\newcommand{\eq}[1]{\begin{equation} #1 \end{equation}}
\newcommand{\al}[1]{\begin{align} #1 \end{align}}
\newcommand{\ml}[1]{\begin{multline} #1 \end{multline}}

\begin{document}
\begin{center}
{\bf{\Large A note on spin chain/string duality} \\
\vspace*{.35cm}
}

\vspace*{1cm}
H. Dimov${}^{\ddag}$, 
R.C. Rashkov${}^{\dagger}$\footnote{e-mail: rash@phys.uni-sofia.bg}

\ \\
${}^{\dagger}$Department of Physics, Sofia University, 1164 Sofia,
Bulgaria 

\ \\

${}^{\ddag}$ Department of Mathematics, University of 
Chemical Technology and Metallurgy, 1756 Sofia, Bulgaria

\end{center}

\vspace*{.8cm}

\begin{abstract}
Recently a significant progress in matching the anomalous dimensions
of certain class of operators in N=4 SYM and rotating strings was
made. 
The correspondence was established mainly  by using of Bethe ansatz
technique applied to the spin $s$ Heisenberg chain model. In a recent
paper Kruczenski (hep-th/0311203) suggested to solve the Heisenberg
model by using of sigma model approach. In this paper we generalize the
solutions obtained by Kruczenski and comment on the dual string
theory. It turns out that our solution is related to a generalized
ansatz for rotating strings which can be reduced to the so called 
Neumann-Rosochatius integrable system. We comment on the spin chain 
sigma model and string solutions and the possibilities for a more
precise formulation of on string/gauge theory correspondence.
\end{abstract}

\vspace*{.8cm}

\sect{Introduction}

Recently an important proposal by Berenstein, Maldacena and Nastase \cite{bmn}
made the first step beyond the supergravity approximation in AdS/CFT 
correspondence. The same authors showed in \cite{bmn} how certain operators 
in SYM theory are directly related to string theory in pp-wave limit. 
It was suggested that a certain class of "nearly" chiral operators are 
described by string theory in pp-wave background and vice versa. 
%%%%%%%%%%%
Shortly after this results was given, a proposal for more general
treatment was made. Namely, since the BMN proposal concerns
states with large energy and R-charge, it was realized that in this 
particular limit one can find reliable results using semiclassical 
approximation of string theory around solitonic string configurations 
executing highly symmetric motion in the target space-time. In this 
case the corresponding gauge theory operators are not supposed to be
nearly chiral but they are generically non-chiral \cite{gkp}. This
make the method of rotating strings even more involving. While on
string theory side many rotating strings solutions in $AdS_5\times
S^5$ and other \cite{ts}, less supersymmetric backgrounds were quickly found
\cite{rash}, the relation to the gauge theory is still not given in
such a direct way as in the case of string theory in pp-wave background. 
Nevertheless, some important steps were made. It was observed by
Minahan and Zarembo \cite{mz} that SYM theory can be mapped to spin 
chain integrable system. This turns out to be great advantage because 
following \cite{fadd}, one can diagonalize the matrix of the anomalous 
dimensions by using the algebraic Bethe ansatz technique. More
specifically, let us consider operators in $\N =4$ SYM theory. The R-symmetry
group is $SO(6)$ whose bosonic sector contains six real scalar fields 
in the adjoint, $\phi^A$, $A=1, ...,6$, which can be organize in three 
complex scalar fields, $X=\phi^1+i\phi^2, Y=\f^3+i\f^4,
Z=\f^5+i\f^6$. The operators in question are of the form
$tr(X^{J_1}Y^{J_2}Z^{J_3})$ where $J_i$ are large. For instance,
significant progress in pp-wave string/SYM correspondence was made in
the sector of operators with large number of $Z$'s and two other
scalars, considered as impurities\footnote{for three impurities see also \cite{khoze}},
\eq{
{\mathcal O}\sim tr(XYZ\dots Z).
\label{i1}
}
We note that in rotating strings approach \cite{gkp,ts1,mz} the form
of the operators (\ref{i1}) is more general as mentioned above, namely
\eq{
{\mathcal O}\sim tr(XZ\dots YY\dots Z\dots Z)
\label{i2}
}

In both approaches in order to obtain the anomalous dimensions one must
diagonalize the action of the dilatation operator acting on these
operators. One loop dilatation operator, for instance, has the form \cite{beis,stau}
\eq{
D_2=\l\sum\limits_{i=1}^J(\frac 14-\vec S_i\vec S_{i+1}),
\label{i3}
}
i.e., it acts as permutation operator between near neighborhood
operators. One can then map the operators of the form (\ref{i2}) into 
Heisenberg spin chain system which is known to be integrable. For
operators of two species, $X$ and $Z$ for instance there exists the identification
\eq{
ZZX\dots ZXX\Leftrightarrow
|\uparrow\uparrow\downarrow\dots\uparrow\downarrow
\downarrow\rangle,
\label{i4}
}
i.e., we identify the $Z$ fields with spin up and $X$ fields with spin
down. A powerful tool in studying this integrable system is Bethe
ansatz technique which was used to this particular problem first in
\cite{mz}. It was observed later \cite{tsstau} that the anomalous
dimension computed in spin chain model in the first loop approximation
and those obtained from rotating strings agrees exactly. This suggests
that one can map (non-locally perhaps) the integrable spin chain model
to the string sigma model \cite{ts2}. An important step toward the
clarification of the hypothesis that the continuous spin chain
action can be mapped directly to the string sigma model action was
recently made by Kruczenski \cite{krucz}. The starting point is
the Heisenberg spin chain in its ferromagnetic phase. Taking the
length of the spin chain $J$ to be very large one obtains continuous 
ferromagnetic sigma model. It was argued that the resulting action coincides with
rotating string sigma model in certain limit and a consistency check
for a simple solution to the spin chain and rotating string sigma
model was given.

In this short note we consider more general solution to the spin chain 
sigma model. In our case it turns out that the corresponding rotating 
string ansatz is not that of Neumann integrable system as in
\cite{krucz}\footnote{For more details see also \cite{tsneum}} but of
so called Neumann-Rosochatius integrable system \cite{tsnr}. We
analyze the solutions to these systems and find out that, up to one loop,
they agrees exactly. Although one can expect that there should be
some rotating string configuration that corresponds to the more
general solutions of the spin chain model, it seems that our result
gives a non-trivial check of the conjecture made in \cite{krucz}. It shows
also that the map between the two models is not that straightforward
as it might seems at first glance.

In the next section we give a brief review of the large spin chain
length limit which corresponds to the continuous limit of Heisenberg spin
chain sigma model and the particular limit of the Polyakov string
sigma model as proposed in \cite{krucz}. In next section we find a more
general solution to the spin chain model. The corresponding rotating
string ansatz is found and discussed. In the concluding section we
comment on our results and some open problems.

\sect{Review of spin chain/strings duality}

In this section we give a brief review of the continuous spin chain
sigma model and its relation to string sigma model. It is well known
that the Heisenberg model is related to the gauge theories in the
large N limit. By making use of holographic correspondence, the
relation between spin chains and string theory was recently studied 
\cite{stau,mz}. These authors successfully used the algebraic Bethe
ansatz (see for review \cite{fadd}) to obtain perfect matching of the 
anomalous dimensions (in one loop) obtained from string theory (the
calculations are actually valid to all loops) and those obtained from
gauge theory side. In the same time, however, there are string
solutions with very complicated relations between the energy and spins 
(see for instance \cite{frolts}) which make the calculation of the
anomalous dimensions very hard. To obtain the corresponding gauge
theory counterpart, it is desirable to have more direct way to
establish the correspondence between the two theories. 

Recently Kruczenski suggested such a correspondence based on the
equivalence of the continuous ferromagnetic sigma model (which is
equivalent to spin $s$ Heisenberg spin chain) and string sigma model
in specific limit \cite{krucz}. In this section we will follow that
line of presentation.

We start with the realization of the spin chain sigma model. The
easiest way is may be to use the coherent states on the target space 
$SU(2)/U(1)=S^2$ \cite{frad,perel}, i.e. by using $\vec n$ field 
formalism
\eq{
|\vec n\rangle=e^{iS_z\phi}e^{iS_y\theta}|ss\rangle .
\label{2.1}
}
We will use the standard parametrization of $S^2$
\eq{
\vec n=(\sin\theta\cos\phi,\sin\theta\sin\phi,\cos\theta)
\label{2.2}
}
\eq{
S_z|ss\rangle=s|ss\rangle,
\label{2.3}
}
and
\eq{
\vec n^2=1 .
\label{2.4}
}
Starting from the standard lagrangian
\eq{
\L=\frac{s}{4}(\p_\m\vec n)^2
\label{2.5}
}
and introducing canonical conjugate variable
\eq{
\vec p=\frac{s}{2}\p_t\vec n\times\vec n
\label{2.6}
}
one can easily deduce the canonical Poisson brackets
\al{
& \{p^a,p^b\}=\varepsilon^{abc}p^c\d(x-x') \notag \\
& \{p^a,n^b\}=\varepsilon^{abc}n^c\d(x-x') \label{2.7} \\
& \{n^a,n^b\}=0.
}
To make contact with the spin chain, we regularize the model as follows
\eq{
\vec p_k=\Delta\vec p(\s);\quad \vec n_k=\vec n(\s);\quad \s=k\Delta.
\notag
}
Then the sigma model variables in large spin limit can be realized by
a pair of spin variables $S_{2k-1},S_{2k}$ as follows
\al{
& p_k=S_{2k-1}+S_{2k} \notag \\
& n_k=\frac{1}{2s}(S_{2k}-S_{2k-1})
\label{2.8}
}
On the other hand,in order to obtain the sigma model effective action one can
follow \cite{frad} and use path integral. The result is
\eq{
S(\vec n)=s\sum\limits_k\int dt\int\limits_0^1d\t(\vec n_k.\p_\m\vec 
n_k\times\p_\n\vec n\varepsilon^{\m\n}) -\frac{\tilde \l s^2}{2}\int 
dt\sum\limits_k|\vec n_k-\vec n_{k+1}|^2
\label{2.9}
}
We are interested, however, in the limit of large spin J, which is
actually a long wave limit. In this limit the variable $\s$ becomes
continuous and is running from $0$ to $J$. By making use of the
explicit parametrization of the $\vec n$ field and the standard
formula $\Delta\sum_k=\int d\s$, one can obtain the following
effective action
\eq{
S(\vec n)=s\int dtd\s\int\limits_0^1d\t \sin\theta\p_\m\phi
\times\p_\n\theta\varepsilon^{\m\n}-\frac{\tilde \l s^2}{2} \int d\s 
dt[(\p_\s\theta)^2+\sin^2\theta(\p_\s\phi)^2].
\label{2.10}
}
We will not go into details here but only point out that the ground
state of the system is ferromagnetic and only long range fluctuations 
take place.

We are actually interested in the conserved quantities, energy and
spin, which define the anomalous dimensions in the gauge theory. These
are given by
\al{
& S_z=P_\phi=-s\int d\s\int d\t\sin\theta\p_\t\theta=-s\int d\s
  \cos\theta 
\label{2.11}\\
& J=\int\limits_0^J d\s \label{2.12} \\
& H=\frac{\tilde \l s^2}{2}\int d\s[(\p_\s\theta)^2+\sin^2\theta(\p_\s\phi)^2].
\label{2.13}
}
To bring the action into a form adapted to our analysis we use that
the volume two-form $\omega_2$ is exact and integrate the Wess-Zumino term
by part (excluding the poles of the sphere where the usual singularity
appears; since our analysis is at classical level it is insignificant)
to obtain
\eq{
S(\vec n)=-s\int d\s dt\cos\theta\p_t\phi -\frac{\tilde \l s^2}{2}
\int d\s dt [(\p_\s\theta)^2+\sin^2\theta(\p_\s\phi)^2].
\label{2.14}
}
From this action one can read off the equations of motion
\al{
& \sin\theta\p_t\theta+\tilde\l s\p_\s(\sin^2\theta\p_\s\phi)=0 
\label{2.15} \\
& \sin\theta\p_t\phi+\tilde\l s\p_\s^2\theta-\tilde\l s
\sin\theta\cos\theta(\p_\s\phi)^2 =0 \label{2.16}
}
supplemented with the boundary conditions
\eq{
\phi(\s=J,t)=\phi(\s=0,t),\quad \theta(\s=J,t)=\theta(\s=0,t).
\label{2.17}
}
In \cite{krucz} the spins and the energy for a particular ansatz were
calculated. It was demonstrated that the energy E and the ratio
$J_2/J$ for this particular solution exactly coincide with those
obtained by using of Bethe ansatz results \cite{stau}. On the other hand, in 
\cite{frolts} it was proven that the Bethe ansatz results and string
theory calculations exactly agrees in one loop.  To be more specific, 
the ansatz used in \cite{krucz} to solve the equations of motion is
\eq{
\p_\s\phi=0 \notag
}
which, by using (\ref{2.15}), immediately implies that $\p_t\theta=0$ 
and therefore 
\eq{
\p_t\phi=\omega. \notag
} 
With this ansatz the equations of motion (\ref{2.15})and (\ref{2.16})
are greatly simplified and the conserved quantities defining the anomalous 
dimensions can be easily calculated 
\al{
& S_z=-s\sqrt{\frac{2}{b}}\{2E(x)-K(x)\} \notag \\
& J=\frac{a+b}{2b}, \qquad x=\frac{a+b}{2b} \notag
} 
and
\al{
& \frac{J_2}{J}=(s+\frac{1}{2})-2s\frac{E(x)}{K(x)} \notag \\
& E=\frac{\tilde\l}{J}32 s^2 K(x)[E(x)-(1-x)K(x)] \notag
}
which is exactly the same as the one loop result from rotating string
in \cite{frolts}\footnote{For more details see \cite{krucz}.}.

We will focus now on the string sigma model. An interesting
observation is that the spin chain picture looks like rotating string
but with no motion of the center of mass (which actually is a feature
of pp-wave limit). This is an useful instruction to look for rotating
string solution in pp-wave like metric. We are looking for two spin
solution in $S^5$ part of $AdS_5\times S^5$ background which means
that one should take $S^3\subset S^5$. The relevant part of the metric
defining string sigma model is
\eq{
ds^2=-dt^2+d\psi^2+\cos^2\psi d\phi_1^2+\sin^2\psi d\phi_2^2.
\label{2.18}
}
The string solution was obtained in \cite{frolts} but in order to
compare the result with spin chain, it is useful to make two steps
change of variables:
\eq{
\phi_1=\varphi_1+\varphi_2, \quad \phi_2=\varphi_1-\varphi_2
\notag
} and
\eq{
\varphi\to t+\varphi_1,
\label{2.19}
}
which yield
\eq{
ds^2=2dt(d\varphi_1+\cos(2\psi)d\varphi_2)+d\psi^2+
d\varphi_1^2+d\varphi_2^2+
2\cos(2\psi)d\varphi_1d\varphi_2
\label{2.20}
}
Taking $t=\kappa\t$ the Polyakov action in the above background 
becomes\footnote{We follow the notations of \cite{krucz}}
\al{
S=
\frac{R^2}{4\pi\a'}\int[2\kappa\dot\varphi_1+\dot\psi^2+\dot\varphi_1^2
  +
& \dot\varphi_2^2+2\cos(2\psi)\kappa\dot\varphi_2+2\cos(2\psi)
\dot\varphi_1\dot\varphi_2 
\notag \\
& -{\psi'}^2-{\varphi'_1}^2-{\varphi'_2}^2-2\cos(2\psi)
\varphi'_1\varphi'_2]
\label{2.21}
}
with the corresponding Virasoro constraints
\ml{
2\kappa\varphi'_1+\dot\psi\psi'+\dot\varphi_1\varphi'_1+
\dot\varphi_2\varphi'_2
+2\cos(2\psi)\kappa\varphi'_2+2\cos(2\psi)\dot\varphi_1\varphi'_2
\\
+2\cos(2\psi)\dot\varphi_2\varphi'_1=0, \label{2.22}
}
\ml{
2\kappa\dot\varphi_1+\dot\psi^2+\dot\varphi_1^2+\dot\varphi_2^2+
2\cos(2\psi)\kappa \dot\varphi_2+2\cos(2\psi)\dot\varphi_1\dot\varphi_2
\\
+2\cos(2\psi)\varphi'_1\varphi'_2
+{\psi'}^2+{\varphi_1'}^2+{\varphi_2'}^2
=0. \label{2.23}
}
A particular class of two spins solutions corresponding to rotating
strings in $S^3\subset S^5$ were studied in details in \cite{frolts}. 
According to \cite{gkp} the results are reliable in the limit of large 
energy and spins, i.e. the limit where the semiclassical approximation 
is valid. Therefore, we are looking for the terms in the Polyakov
action contributing to this particular limit. As it was argued in
\cite{krucz}, the case of large spin and energy corresponds to the
limit $\kappa\to\infty$. The non-trivial contribution actually comes
from the limit
\eq{
\kappa\to\infty, \quad \dot X^\m\to0, \quad \kappa\dot X^\m=fixed
\label{2.24}
}
The terms in (\ref{2.21}), (\ref{2.22}) and (\ref{2.23}) which survive
this limit give the following resulting action and constraints
\ml{
S=\frac{R^2}{4\pi\a'}\int[2\kappa\dot\varphi_1+2\cos(2\psi)\kappa\dot
\varphi_2 -{\psi'}^2-{\varphi_1'}^2-{\varphi_2'}^2-\\
2\cos(2\psi)\varphi_1'\varphi_2']
\label{2.25}
}
and
\al{
& \kappa\varphi_1'+\cos(2\psi)\kappa\varphi_2'=0 \label{2.26}, \\
& 2\kappa\dot\varphi_1+2\cos(2\psi)\kappa\dot\varphi_2+{\psi'}^2 +
{\varphi_1'}^2+{\varphi_2'}^2+2\cos(2\psi)\varphi_1'\varphi_2'=0.
\label{2.27}
}
The field $\varphi_1$ somehow decouples from the system being
completely determined from the constraints (\ref{2.26}) and
(\ref{2.27}). Finding $\varphi_1$ from the constraints and
substituting it in the action (\ref{2.26}) we obtain
\eq{
S=\frac{R^2}{4\pi\a'}\int[2\kappa\dot\varphi_1+2\cos(2\psi)\kappa
\dot\varphi_2 -{\psi'}^2-sin^2(2\psi){\varphi_2'}^2]
\label{2.28}
}
If we compare this action to the action for spin chain (\ref{2.14}) we see
that both are completely equivalent. Actually, they coincide if we
make use of the identification
\eq{
\tilde\s=\frac{R^2}{4\pi\a'}2\kappa\s,\quad \varphi_2=-\frac 12\phi,
\quad \psi=\frac 12\theta \label{2.29}
}
and the well know relation $\l=R^4/{\a'}^2$.

The rotating string solution that corresponds to the spin chain
solution was investigated by Frolov and Tseytlin \cite{frolts}. The
ansatz used there is
\eq{
t=\kappa\t,\quad \phi_1=\omega_1\t,\quad \phi_2=\omega_2\t, 
\quad \theta=\theta(\s)
\label{2.30}
}
with the following expressions for the energy and spins
\eq{
E=\kappa, \quad J_1=\frac{2\omega_1}{\sqrt{\omega_{12}^2}}E(x)
\label{2.31}
}
where $x=(\kappa^2-\omega_1^2)/\omega_{12}^2$ and $E(x)$ is the complete
elliptic integral of second kind. The periodicity condition determines 
$\omega_{12}$ in terms of complete elliptic integral of first kind
\eq{
\sqrt{\omega_{12}^2}=\frac{2}{\pi}K(x)
\label{2.32}
}
We conclude this section reffering for details to \cite{frolts,krucz}.

\sect{A new case of spin chain/string correspondence}

In this section we generalize the solution to the spin chain sigma
model obtained in \cite{krucz} and discuss on the corresponding string
solutions. As mentioned above, the spin $s$ Heisenberg model has the
following equations of motion
\al{
& \dot\theta\sin\theta+\tilde\l s\p_\s[\sin^2\theta\p_\s\phi]=0 \label{3.1} \\
& \dot\phi\sin\theta+\tilde\l s\theta''-\tilde\l 
s\sin\theta\cos\theta(\p_\s\phi)^2=0
\label{3.2}
}
We look for more general solution of the equations of motion
(\ref{3.1}) and (\ref{3.2}) requiring that $\theta$ is function of 
$\s$ only, i.e.
\eq{
\dot\theta=0 \label{3.3}
}
With no other restrictions, one can find that $\phi$ satisfies the 
relation
\eq{
\phi'=\frac{A}{\sin^2\theta}.
\label{3.4}
}
Substituting (\ref{3.4}) in (\ref{3.2}) we obtain
\eq{
\dot\phi\sin\theta+\tilde\l s\theta''-\tilde\l s
\frac{A\cos\theta}{\sin^3\theta}=0
\label{3.5}
}
From (\ref{3.5}) we can find the equation that determines completely 
$\phi$
\al{
& \ddot\phi=0, \notag \\
& \dot\phi'=0 \label{3.6}
}
with the obvious solution
\eq{
\phi=\omega\tau+\hat\phi(\s); \quad \hat\phi'=\frac{A}{\sin^2\theta}.
\label{3.7}
}
Using (\ref{3.7}) we derive the final equation for $\theta$ 
variable
\eq{
\omega\sin\theta+\tilde\l s\theta''-\tilde\l s\frac{A\cos\theta}
{\sin^3\theta}=0
\label{3.8}
}
As a result we have simple equations for $\phi$ and $\theta$ that
determine the invariants of the model - the energy and spins. Let us
concentrate on the solution of (\ref{2.8}). Multiplying (\ref{3.8})
by $\theta'$ and integrating the resulting equation once we get 
\eq{
\lb(\frac{d\theta}{d\s}\rb)^2+\frac{A^2}{\sin^2\theta}-
\frac{2\omega}{\tilde\l s}\cos\theta=B.
\label{3.9}
}
If we denote
\eq{
\x=\cos\theta, \label{3.10}
}
one can rewrite (\ref{3.9}) in the form
\eq{
\lb(\frac{d\x}{d\s}\rb)^2=\a\x^3-B\x^2-\a\x+B-A^2,
\label{3.11}
}
where we define
\eq{
\a=-\frac{2\omega}{\tilde\l s}>0;\quad i.e. \,\, \omega<0 .
\label{3.12}
}
It is useful to define a new variable by
\eq{
\x=\frac{B}{3\a}+x .
\label{3.13}
}
In terms of the new variable $x$ the equation (\ref{3.11}) takes the form
\eq{
\lb(\frac{dx}{d\tilde\s}\rb)^2=4x^3-g_2x-g_3,
\label{3.14}
}
where
\al{
& \tilde\s=\frac{\sqrt{\a}}{2}\s, \quad g_2=4\frac{B+3\a^2}{3\a^2}. \label{3.15} \\
& g_3=\frac{4}{3\a}\lb(3A^2+\frac{2B^3}{9\a^2}-2B\rb). \label{3.16}
}
The equation (\ref{3.14}) defines an elliptic curve in Weierstrass
form. The solution of the equation defines an elliptic integral whose
inverse is Weierstrass or Jacobi elliptic functions. Since we are
looking for finite periodic solutions, the integration constants
should be adjusted so that the solution is Jacobi elliptic
function. Let us denote by $e_1,\,e_2,\,e_3$ the roots of the right
hand side of (\ref{3.14}) which, as it is well known from the theory 
of elliptic functions, satisfy the relations
\al{
& \sum\limits_k e_k=0, \quad e_1e_2e_3=\frac{1}{4}g_3, \label{3.1.17} \\
& e_1e_2+e_2e_3+e_3e_1=-\frac{1}{4}g_2 .\label{3.18}
}
One can write then (\ref{3.14}) in the form
\eq{
\lb(\frac{dx}{d\tilde\s}\rb)^2=4(x-e_1)(x-e_2)(x-e_3).
\label{3.19}
}
It is easy to bring eq.(\ref{3.19}) into Jacobi form by using the 
transfromation
\eq{
x=e_1+e_{21}\e^2 \label{3.20}.
}
Finally, we end up with
\eq{
\lb(\frac{d\e}{d\tilde\s}\rb)^2=e_{31}(1-\e^2)(1-\k\e^2),
\label{3.21}
}
where the modulus $\k$ is given by $\k=e_{21}/e_{31}$, and 
$e_{nm}=e_n-e_m$.
The finite periodic solution to eq. (\ref{3.21}) satisfying $\e(0)=0$ is
\eq{
\e=sn(\sqrt{e_{31}}\tilde\s,\k)=sn\lb(\frac{\sqrt{\a e_{31}}}{2}\s,\k\rb)
\label{3.22}
}
and the expression for $\cos\theta$ becomes
\eq{
\cos\theta=\frac{B}{3\a}+e_1+e_{21}sn^2(\frac{\sqrt{\a e_{31}}}{2}\s,\k)
\label{3.23}
}
In addition we have to satisfy the boundary conditions
\eq{
\theta(\s=J,t)=\theta(\s=0,t)
\notag
}
which determines $\a$ in terms of the complete elliptic integral of 
first kind
\eq{
\frac{\sqrt{\a e_{31}}}{2}J=2K(\k),\quad \text{or} \quad J=\frac{4K(\k)}{\sqrt{\a e_{31}}}
\label{3.24}
}
and
\eq{
\cos\theta=\frac{B}{3\a}+e_1+e_{21}sn^2(\frac{2K(\k)}{J}\s,\k)
\label{3.25}
}
Having the solution (\ref{3.25}) one can find the conserved quantities
as energy and spins. First of all the total spin $J$ is already
(consistently) determined by the periodicity condition
\eq{
J=\frac{4K(\k)}{\sqrt{\a e_{31}}} .
\label{3.26}
}
The calculation of $S_z$ is also straightforward
\al{
 S_z=\frac{J_2-J_1}{2}& =-s\int\limits_0^J\cos\theta d\s=-s
\lb[\frac{B}{3\a}+e_1+ e_{31}(1-\frac{E(\k)}{K(\k)})\rb] \notag \\
& =-s\frac{4\varpi}{\sqrt{\a}}\lb[\frac{B}{3\a}+e_3-e_{31}
\frac{E(\k)}{K(\k)}\rb],
\label{3.27}
}
where $\varpi$ is the primitive period of the Weierstrass $\wp$ function.
The energy can be also easily determined as follows. First, we use
(\ref{3.4}) and (\ref{3.9}) to get
\al{
E&=\frac{\tilde\l s^2}{2}\int\limits_0^J\lb[(\p_s\theta)^2
+\sin^2(\p_s\phi)^2\rb]=
\frac{\tilde\l s^2}{2}\int\limits_0^J\lb[B-\a\cos\theta\rb] \notag \\
& =\frac{\tilde\l s^2}{2}\lb[BJ+\frac{\a}{s}S_z\rb] .
\label{3.28}
}
Using (\ref{3.26}) and (\ref{3.27}) we find for the energy the next 
expression
\eq{
E=\frac{2\tilde\l s^2\varpi}{\a}\lb[\frac{2B}{3}- \a e_3+\a e_{31}E(\k)\rb] .
\label{3.29}
}
The spins $J_i$ are found to be
\eq{
J_1=\frac{2\varpi}{\sqrt{\a}}\lb[1+2s\lb(\frac{B}{3\a}+e_3\rb)-2se_{31}
\frac{E(\k)}{K(\k)}\rb]
\label{3.30}
}
and
\eq{
J_2=\frac{2\varpi}{\sqrt{\a}}\lb[1-2s\lb(\frac{B}{3\a}+e_3\rb)+2se_{31}
\frac{E(\k)}{K(\k)}\rb]
\label{3.31}
}

To summarize, we found explicit expressions for the characteristics of
the spin chain sigma model that determine the anomalous dimensions.

We turn now to the string side. The consideration of the Polyakov
action in the limit $\hat\k\to\infty, \dot X\to0, \hat\k\dot X=fixed$ 
(here we changed the notation $t=\hat\k\tau$) ended up in the previous 
section with the expression
\eq{
S=\frac{R^2}{4\pi\a}\int\lb[\a\hat\k\dot\varphi_1+2\cos(2\psi)\hat\k
\dot\varphi_2 -{\psi'}^2-{\varphi_1'}^2-{\varphi_2'}^2-2\cos(2\psi)
\varphi_1'\varphi_2'\rb]
\label{3.32}
}
with the conformal constraints
\al{
& 2\hat\k\varphi_1'+2\cos(2\psi)\hat\k\varphi_2'=0, \notag \\
& 2\hat\k\dot\varphi_1+2\cos(2\psi)\hat\k\dot\varphi_2+{\psi'}^2+
{\varphi_1'}^2+{\varphi_2'}^2
+2\cos(2\psi)\varphi_1'\varphi_2' .
\label{3.33}
}
Since we found more general solution for the spin chain allowing
$\phi$ to depend on $\s$, we expect that the corresponding string
solutions are also more general, allowing $\s$ dependence for
$\varphi_i$. If so, due to (\ref{3.33}) the variables $\varphi_i$ are
related by
\eq{
\varphi_1'=-\cos(2\psi)\varphi_2 .
\label{3.34}
}
If we substitute (\ref{3.34}) into (\ref{3.32}) we will obtain again
(\ref{2.28}) which after identification $\varphi_2=-\phi/2,
\psi=\theta/2$ and $\tilde\s=R^22\hat\k\s/4\pi\a'$ coincides again
with the action for the spin chain model. Assuming that $\theta$ is $\t$ 
independent, a natural ansatz for $\varphi_i$ is
\al{
& \varphi_1=\omega_1\t+\a_1(\s) \notag \\
& \varphi_2=\omega_2\t+\a_2(\s) .
\label{3.35}
}
In contrast to the ansatz (\ref{2.30}) in the previous section, which
actually reduces the problem to the Neumann integrable system
\cite{tsneum}, the ansatz (\ref{3.35}) is related to the so called
Neumann-Rosochatius (N-R) integrable system \cite{tsnr}. Hence, we
are to consider two spin string solution of N-R system (we are
confined on $S^3\subset S^5$ and therefore we can have at most two
spins $J_2$ and $J_2$). 

The most general ansatz studied in \cite{tsnr} is
\eq{
\phi_i=\omega_i\t+\a_i(\s), \,\, i=1,2,3;\quad \psi=\psi(\s), 
\,\, \g=\g(\s)
\label{3.36}
}
where the angles (\ref{3.36}) parameterize $S^5$ as follows
\al{
& X_1+iX_2=\sin\g\cos\psi e^{i\phi_1}=z_1(\s)e^{\omega_1\t}, \notag \\
& X_3+iX_4=\sin\g\sin\psi e^{i\phi_2}=z_2(\s)e^{\omega_2\t}, \label{3.37} \\
& X_5+iX_6=\cos\g e^{i\phi_3}=z_3(\s)e^{\omega_3\t} .\notag 
}
It is useful to introduce variables $r_i(\s)$ by
\eq{
z_i(\s)=r_i(\s)e^{i\a_i(\s)}, \quad i=1,2.\label{3.38} 
}
In our case we have $\g=\pi/2$ and $\a_3=0$ which means that
\al{
& r_1(\s)=\cos\psi, \notag \\
& r_2(\s)=\sin\psi, \quad r_3=0 .
\label{3.39}
}
The string lagrangian for the relevant variables is
\al{
& \L=\frac{1}{2}\sum(z_i'{z_i^\star}'-\omega_i^2z_iz_i^\star) +
\frac{1}{2}\Lambda(\sum 
z_iz_i^\star-1) \notag \\
& =\frac{1}{2}\sum({r_i'}^2+r_i^2{\a_i'}^2-\omega_i^2r_i^2)+
\frac{1}{2}\Lambda(\sum r_i^2-1) .
\label{3.40}
}
The equation of motion for $\a_i$ gives
\eq{
\a_i'=\frac{v_i}{r_i^2}, \quad i=1,2.
\label{3.41}
}
Substitution of (\ref{3.41}) into (\ref{3.40}) leads to the so called
Neumann-Rosochatius integrable system defined by the lagrangian
\eq{
\L=\frac{1}{2}\sum\lb[{r_i'}^2+\frac{v_i^2}{r_i^2}-\omega_i^2r_i^2\rb]+
\frac{1}{2}\Lambda(\sum r_i^2-1) .
\label{3.42}
}
The solution of N-R system was studied in \cite{tsnr} by using
elliptic coordinates. For our purpose, however, it is more convenient to
study the system in global coordinates.

The equations of motion derived from (\ref{3.42})
\eq{
r_i''=-\omega_i^2r_i-\frac{v_i^2}{r_i^3}-r_i\sum\lb({r_j'}^2-\omega_i^2r_j^2
+\frac{v_j^2}{r_j^2}\rb)
\label{343}
}
can be written in global coordinates using the relations
(\ref{3.39}). Multiplying the first equation in (\ref{3.42}) by $r_2$
and subtracting the second one multiplied by $r_1$, one finds
\eq{
\psi''+\omega_{21}^2\sin\psi\cos\psi+\frac{v_2^2\cos\psi}{\sin^3\psi}- 
\frac{v_1^2\sin\psi}{\cos^3\psi}=0,
\label{3.44}
}
where
\eq{
\omega_{21}^2=\omega_2^2-\omega_1^2.
\label{3.45}
}
One can integrate (\ref{3.44}) once to obtain
\eq{
{\psi'}^2+\omega_{21}^2\sin^2\psi-\frac{v_2^2}{\sin^2\psi}-
\frac{v_1^2}{\cos^2\psi}=\hat A,
\label{3.46}
}
or, equivalently
\eq{
\lb(\sin\psi\cos\psi\frac{d\psi}{d\s}\rb)^2=
v_2^2+(\hat A-v_{21}^2)\sin^2\psi-(\hat A+\omega_{21}^2)\sin^4\psi+
\omega_{21}^2\sin^6\psi,
\label{3.47}
}
where $v_{21}^2=v_2^2-v_1^2$. Defining the variable
\eq{
\x_{str}=\sin^2\psi
\label{3.48}
}
we find the equation of motion as an elliptic curve equation in
Weierstrass form
\eq{
\lb(\frac{d\x_{str}}{d\s}\rb)^2=4[\omega_{21}^2\x_{str}^3-(\omega_{21}^2+
\hat A)\x_{str}^2- (v_{21}^2-\hat A)\x_{str}+v_2^2].
\label{3.49}
}
One can bring the equation (\ref{3.49}) into Weierstrass from by
defining a new variable 
\eq{
\x_{str}=\frac{\omega_{21}^2+\hat A}{3\omega_{21}^2}+\tilde x .
\label{3.50}
}
The equation for $\tilde x$ then becomes
\eq{
\lb(\frac{d\tilde x}{d\tilde\s}\rb)^2=4\tilde x^3-\tilde g_2
\tilde x-\tilde g_3
\label{3.51}
}
where
\al{
& \tilde\s=\sqrt{\omega_{21}^2}\s \label{3.52} \\
& \tilde g_2=\frac{4}{\omega_{21}^2}\lb(v_{21}^2+
\frac{\omega_{21}^2+\hat A}{3\omega_{21}^2}-\hat A\rb), \label{3.53} \\
& \tilde g_3=\frac{4}{\omega_{21}^2}\lb[\frac{(v_{21}^2-\hat
    A)(\omega_{21}^2+\hat A)}
{3\omega_{21}^2} +\frac{2(\omega_{21}^2+\hat A)^3}{27\omega_{21}^4}-v_2^2\rb].
\label{3.54}
}
Denoting the roots of the right hand side of (\ref{3.51}) by $\tilde
e_1,\tilde e_2,\tilde e_3$ we get
\eq{
\lb(\frac{d\tilde x}{d\tilde\s}\rb)^2=4(\tilde x-\tilde e_1)
(\tilde x-\tilde e_2) (\tilde x-\tilde e_3)
\label{3.55}
}
with the same relations as in (\ref{3.21}), (\ref{3.22}) but now
$\tilde g_i$ are given by (\ref{3.53}) and (\ref{3.54}). Using the 
transformation
\eq{
\tilde x=\tilde e_1+\tilde e_{21}\tilde\e^2
\label{3.56}
}
we get
\eq{
\lb(\frac{d\tilde\e}{d(\sqrt{\tilde e_{31}}\tilde\s}\rb)^2=
(1-\tilde\e)(1-\tilde\k\tilde\e),
\label{3.57}
}
where
\eq{
\tilde\k=\frac{\tilde e_{21}}{\tilde e_{31}}.
\label{3.58}
}
We note that the equation for $\e$ in spin chain case has exactly the
same form as the equation (\ref{3.57}) for $\tilde\e$ in the string 
case.

The solution of (\ref{3.57}) is
\eq{
\tilde\e=sn(\sqrt{\omega_{21}^2\tilde e_{31}}\s,\tilde\k).
\label{3.59}
}
and the solution for the original variable $\sin^2\psi$ is given by
\eq{
\sin^2\psi=\frac{\omega_{21}^2+\hat A}{3\omega_{21}^2}+\tilde e_1+\tilde e_{21}
sn^2\lb(\sqrt{\omega_{21}^2\tilde e_{31}}\s,\tilde\k\rb).
\label{3.60}
}
The periodicity condition requires
\eq{
\sqrt{\omega_{21}^2\tilde e_{31}}=\frac{2K(\k)}{\pi} .
\label{3.61}
}
Now one can find the spins $J_1$ and $J_2$
\eq{
J_1=\omega_1\lb(1-\frac{\omega_{21}^2+\hat A}{3\omega_{21}^2}-\tilde e_3+
\tilde e_{31}\frac{E(\k)}{K(\k)}\rb)
\label{3.62}
}
and
\eq{
J_2=\omega_2\lb(\frac{\omega_{21}^2+\hat A}{3\omega_{21}^2}+\tilde e_3-
\tilde e_{31}\frac{E(\k)}{K(\k)}\rb).
\label{3.63}
}
As expected, the form of the spins $J_1$ and $J_2$ in spin chain and string
cases are similar but not exactly the same. As we discussed in Section
2 the correspondence is valid only in one loop. To make the agreement
exact, let us evaluate the terms which we have neglected in the action
(\ref{2.21}). First of all, we made change of variables as follows
\al{
& t=\k\t; \quad \f_i=\omega_i\t+\a(\s), \notag \\
& \f_1=\varphi_1+\varphi_2+t; \quad \f_1=\varphi_1-\varphi_2+t.
\label{3.64}
}
The terms in the action which are small then become
\eq{
\Delta S=\frac{R^2}{4\pi\a'}\int \lb[\dot\psi^2+\dot\varphi_1^2+
\dot\varphi_2^2+
2\cos(2\psi)\dot\varphi_1\dot\varphi_2\rb] .
\label{3.65}
}
Using the ansatz (\ref{3.64}) we get
\al{
&\dot\varphi_1=-\frac{(\k-\omega_1)+(\k-\omega_2)}{2}=
-(\epsilon_1+\epsilon_2), \notag \\
&\dot\varphi_1=-\frac{(\k-\omega_1)-(\k-\omega_2)}{2}=
-(\epsilon_1-\epsilon_2), \notag \\
\label{3.66}
}
where we set $\epsilon_i=(\k-\omega_i)/2$. The expression (\ref{3.65})
then becomes
\al{
\Delta S&=\frac{R^2}{4\pi\a'}\int \lb[\varepsilon_1+
\varepsilon_2+2\cos(2\psi)\varepsilon_1\varepsilon_2 \rb]\notag \\
&=\frac{R^2}{4\pi\a'}\lb[\varepsilon_1^2\varepsilon_2^2+ 2
\frac{\varepsilon_1\varepsilon_2}{\omega_1\omega_2}\lb(J_2\omega_1-
J_1\omega_2\rb)\rb],
\label{3.67}
}
where
$\varepsilon_1=\epsilon_1+\epsilon_2,\,\,\varepsilon_2=\epsilon_1 
-\epsilon_2$. Using the explicit expressions for the spins
(\ref{3.62}, \,\,\ref{3.63}) we find
\eq{
\Delta S=\frac{R^2}{2\a'}\lb[(\varepsilon_1+\varepsilon_2)^2- 
4\varepsilon_1\varepsilon_2 (a-\tilde e_{31}
\frac{E(\tilde\k)}{K(\tilde\k)})\rb],
\label{3.68}
}
where $a=(\hat\omega_{21}^2+\hat A)/3\hat\omega_{21}^2+\tilde e_3$.
From (\ref{3.67}) one can conclude that $\varepsilon_i$ are of order 
$\textit{o}(\a'/R^2)$. Since
\al{
& \omega_1=\k-(\varepsilon_1+\varepsilon_2), \notag \\
& \omega_1=\k-(\varepsilon_1-\varepsilon_2),
\label{3.69}
}
the expression for $\omega_1$ and $\omega_2$ in one loop can be written as
\al{
& \omega_1=\k-\textit{o}\lb(\frac{\a'}{R^2}\rb), \notag \\
& \omega_1=\k-\textit{o}\lb(\frac{\a'}{R^2}\rb).
\label{3.70}
}
Redefining the integration constants so that the following equality holds
\eq{
a=\frac 12 -\lb(\frac{B}{3\a}+\tilde e_3\rb),
\label{3.71}
}
we find exact agreement between the solutions in spin chain model and 
rotating strings in one loop.

\sect{Conclusions}

In this paper we considered a generalization of the solutions in spin
chain/string duality proposed in hep-th/0311203 by Kruczenski. In
section 2 we reviewed the proposed spin chain/string duality in the
case of rotating string ansatz elaborated in \cite{tsneum} leading to
Neumann integrable system. In the next section we consider
generalization to the solutions obtained in \cite{krucz} using the
ansatz (\ref{3.3})
\eq{
\dot\theta=0, \quad  \phi=\omega\tau+\hat\phi(\s).
\notag
}
We found the solutions for the spin chain model and the corresponding
enegy and spins. Since the ansatz for the dynamical variables in this
case are of more general form we suggested that the string solutions
in this case should correspond to the more general string ansatz (\ref{3.35})
leading to the Neumann-Rosochatius integrable system
\al{
& \varphi_1=\omega_1\t+\a_1(\s) \notag \\
& \varphi_2=\omega_2\t+\a_2(\s) .
\notag
}
We calculated the conserved quantities in the case of two-spin solutions and found
that the solutions are not exactly of the same form. To make the
correspondence precise we consider in detais the terms that are
neglected when we calculated the string action in a certain limit used
in \cite{krucz}. The calculations show that, after appropriate
redefinitions of the integration constants in the solutions, the
expressions for spins and energy agrees exactly. 

Athough it is expected that the new solutions should be described in
terms of spin chain model, it is interesting, and important, that we
found exact agreement between the two models. It shows that this
approach is valid for the most general known ansatz for rotating
string and the result gives a
strong support of the idea that there should be more direct way to
establish the AdS/CFT correspondence of string theory in $AdS_5\times
S^5$ background.

We note that it would be interesting to apply the same line of 
considerations to the Inozemtsev long range spin chain as discussed in 
\cite{serbstau}.

{\it Note added}: After this paper was completed an interesting paper
\cite{kruryts} appeared. In that paper it was proved that the results
from spin chain and rotating strings agrees to the next order and
a systematic procedure for computing the higher orders in
large $J$ expansion was suggested.

\vspace*{.8cm}

{{\large{\bf Acknowledgements:}} R.R. would like to thank A. Tseytlin
  for comments, email correspondence and critically reading the draft
  of the paper.

\end{document}